\documentclass[seceq]{ptptex}

\usepackage{graphicx}

\title{%        %You can use \\ for explicit line-break.
Percolation Thresholds of the Fortuin-Kasteleyn Cluster
 for a Potts Gauge Glass Model
 on Complex Networks%
}

\subtitle{Analytical Results on the Nishimori Line}    %Use this when you want a subtitle.

\author{%       %Use \scshape for the family name.
Chiaki \textsc{Yamaguchi}%
}

\inst{%     %Affiliation, neglected when [addenda] or [errata].
Kosugichou 1-359, Kawasaki 211-0063, Japan
}

\abst{It was pointed out by de Arcangelis et al. 
 [Europhys.\ Lett.\ \textbf{14} (1991), 515]
 that
 the correct understanding of the percolation phenomenon
 of the Fortuin-Kasteleyn cluster in the Edwards-Anderson model
 is important since a dynamical transition,
 which is characterized by a parameter
 called the Hamming distance or damage, and
 the percolation transition are related to
 a transition for a signal propagating between spins.
 We show analytically the percolation thresholds
 of the Fortuin-Kasteleyn cluster for a Potts gauge glass model,
 which is an extended model of the Edwards-Anderson model, 
 on random graphs with arbitary degree distributions.
 The results are shown on the Nishimori line.
 We also show the results for the infinite-range model. }

\begin{document}
\maketitle

%%%%%%%%%%%%%%%%%%
\section{Introduction}
%%%%%%%%%%%%%%%%%%

The study of spin models on complex networks
 has been carried out \cite{DGM1,IT,KAI}.
 We study a spin model on  
 random graphs with arbitary degree distributions
 as an example of the study of a spin model on complex networks.
 The behavior of spins on a no growing network is investigated.

We investigate a Potts gauge glass model \cite{NS}
 as a spin model.
 The Potts gauge glass model is a spin-glass model
 and is an extended model of the Edwards-Anderson model \cite{EA}
 which is known as a spin-glass model.
 The understanding of the Edwards-Anderson model
 on random graphs and on the Bethe lattice is still incompleted \cite{DGM1, VB, MP}.
 The understanding of the Potts gauge glass model
 on random graphs and on the Bethe lattice is also incompleted.

 The Nishimori line \cite{NS} is a line on the phase diagram
 for the exchange interactions and the temperature.
 The internal energy and the upper bound of the specific heat
 are exactly calculated on the Nishimori line \cite{NS}.
 The location of the multicritical point 
 on the square lattice was conjectured,
 and it was shown that 
 the conjectured value is in good agreement
 with the other numerical estimates 
 \cite{NN}.
 We will obtain results on the Nishimori line.

There is a case where
 a percolation transition of networks occures.
 A network is divided into many networks
 by deleting some of its nodes and/or links.
 There is also a case where a percolation transition
 of clusters occurs. A cluster consists of fictitious bonds.
 The bond is put between spins.
 One of the clusters becomes a giant component when a cluster is percolated.
 We discuss the percolation transition of a cluster.

We investigate
 the percolation transition of the Fortuin-Kasteleyn (FK)
 cluster in the FK random cluster model \cite{FK, KF}.
 In the ferromagnetic spin model,
 the percolation transition point
 of the FK cluster
 agrees with the phase transition point.
 For example,
 the agreement in the ferromagnetic Ising model is described
 in Ref.~\citen{CK}. 
 On the other hand,
 in the Edwards-Anderson model that has a conflict
 in the interactions,
 the percolation transition point of the FK cluster 
 disagrees with the phase transition point.
 It was pointed out by de Arcangelis et al. that,
 despite the disagreement,
 the correct understanding of the percolation phenomenon
 of the FK cluster in the Edwards-Anderson model
 is important since a dynamical transition,
 which is characterized by a parameter
 called the Hamming distance or damage, is
 occurred at a temperature very close to the percolation
 temperature, and the dynamical transition and
 the percolation transition are related to
 a transition for a signal propagating between spins \cite{ACP}.
 We analytically obtain
 the percolation thresholds of the FK cluster 
 for the Potts gauge glass model.

We use a gauge transformation for deriving results.
 The gauge transformation was proposed in Ref.~\citen{NS}.
 In addition to the application of the gauge transformation,
 results are shown by applying a criterion \cite{Y}
 for spin models on the random graphs
 with arbitary degree distributions.

In Ref.~\citen{Y}, by applying the criterion with
 a gauge transformation,
 the percolation thresholds of the FK cluster
 for the Edwards-Anderson model on the random graphs
 with arbitary degree distributions were analytically
 calculated on the Nishimori line.

We also show the results for the infinite-range model.

This article is organized as follows.
 First in \S\ref{sec:2}, a complex network model and the Potts gauge glass model
 are described. The FK cluster is described in \S\ref{sec:3}
 and appendix.
 A criterion for percolation of cluster is
 explained in \S\ref{sec:4}.
 We will find in \S\ref{sec:5} the percolation thresholds.
 This article is summarized in \S\ref{sec:6}.

%%%%%%%%%%%%%%%%%%
\section{A complex network model and a Potts gauge glass model} \label{sec:2}
%%%%%%%%%%%%%%%%%%

A network consists of nodes and links.
 A link connected between nodes.
 The complex network model that
 we investigate is
 random graphs with arbitary degree distributions.
 The networks have no correlation between nodes.
 The node degree, $k$, is given with
 a distribution $p (k)$. 
 The links are randomly connected between nodes.

We define a variable $b (i, j)$, where
 $b (i, j)$ gives one when nodes $i$ and $j$
 is connected by a link.
 $b (i, j)$ gives zero when
 nodes $i$ and $j$ are not connected by the link.
 The degree $k (i)$ of node $i$ is given by
\begin{equation}
 k (i) = \sum_j b (i, j) \, .
\end{equation}
The coordination number (the average of the node degree for links), 
 $\langle k \rangle_N$, is given by 
\begin{equation}
 \langle k \rangle_N = \frac{1}{N} \sum_i^N k (i) \, ,
\end{equation}
 where $\langle \, \rangle_N$ is the average over the entire network.
 $N$ is the number of nodes.
 The average of the square of the node degree for links, 
 $\langle k^2 \rangle_N$, is given by
\begin{equation}
 \langle k^2 \rangle_N = \frac{1}{N} \sum_i^N k^2 (i) \, .
\end{equation}

We define 
\begin{equation}
 a = \frac{2 \langle k \rangle_N}{\langle k^2 \rangle_N} \, , \label{eq:a}
\end{equation}
 where $a$ represents an aspect of the network.

\begin{figure}[t]
\begin{center}
\includegraphics[width=0.95\linewidth]{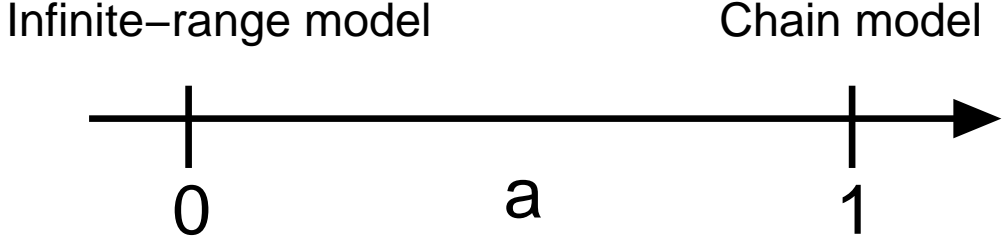}
\end{center}
\caption{
Relation between the aspect $a$ and the model on the network.
\label{fig:a}
}
\end{figure}

Figure~\ref{fig:a} shows the
 relation between the aspect $a$ and the model on the network.
 The network is almost a complete graph when $a$
 is close to zero, and
 the model on the network is almost an infinite-range model.
 The model on the network is the infinite-range model 
 when $\langle k \rangle_N = N - 1$,
 $\langle k^2 \rangle_N = (N - 1)^2$, and $a = 2 / (N - 1)$.
 The network consists of many cycle graphs
 when $a$ is one and $\langle k \rangle_N$ is two.
 The model on the network consists of many chain models
 when $a$ is one and $\langle k \rangle_N$ is two.
 In the Erd\H{o}s-R\'enyi (ER) random graph model and in the Gilbert model,
 the distribution of node degree is the Poisson distribution \cite{DGM1}.
 The ER random graph model is a network model wherein the network
 consists of a fixed number of nodes and
 a fixed number of links, and the links are randomly connected between the nodes.
 The Gilber model is a network model wherein
 the link between nodes is connected with a given probability.
 In the ER random graph model and in the Gilbert model,
 $\langle k^2 \rangle_N = \langle k \rangle_N (\langle k \rangle_N + 1)$,
 and $a = 2 / (\langle k \rangle_N + 1)$.

The Hamiltonian for a Potts gauge glass model, ${\cal H}$, 
 is given by \cite{NS}
\begin{equation}
 {\cal H} = - \frac{J}{2 q}
 \sum^N_i \sum_{\{ j | b(i, j) = 1\}} 
 \sum_{r_{i, j} = 1}^{q - 1} e^{\frac{2 \pi i}{q} ( \nu_{i, j} + q_i - q_j) r_{i, j}} \, ,
 \label{eq:Hamiltonian}
\end{equation}
 where $q_i$ denotes the state of the spin on node $i$, and $q_i = 0, 1, \ldots, q - 1$.
 $\nu_{i, j}$ denotes a variable related to the strength
 of the exchange interaction between
 the spins on nodes $i$ and $j$, and $\nu_{i, j} = 0, 1, \ldots, q - 1$.
 $q$ is the total number of states that a spin takes.

We use representations: $\lambda_i = e^{\frac{2 \pi i}{q} q_i}$ and 
 $J^{(r_{i, j})}_{i, j} = J e^{\frac{2 \pi i}{q} \nu_{i, j} r_{i, j}}$.
 Then, the Hamiltonian (Eq.~(\ref{eq:Hamiltonian}))
 is given by
\begin{equation}
 {\cal H} = -  \frac{1}{2 q} \sum^N_i \sum_{\{ j | b(i, j) = 1\}} 
 \sum_{r_{i, j} = 1}^{q - 1} J^{( r_{i, j} )}_{i, j}
 \lambda^{r_{i, j}}_i
 \lambda^{q - r_{i, j}}_j \, .
\end{equation}

The value of $\nu_{i, j}$ is given with a distribution $P (\nu_{i, j} )$.
 The distribution $P ( \nu_{i,j} )$
 is given by
\begin{equation}
 P ( \nu_{i, j} ) = p \, \delta_{ \nu_{i, j}, 0} + \frac{1 - p}{q - 1}
 (1 - \delta_{ \nu_{i, j}, 0} ) \, , \label{eq:Pnuij}
\end{equation}
 where 
 $p$ is the probability that the exchange interaction between the spins 
 is ferromagnetic. $\delta$ is the Kronecker delta.
 The normalization of $P ( \nu_{i, j} )$ is given by
\begin{equation}
 \sum^{q - 1}_{\nu_{i, j} = 0}
 P ( \nu_{i, j} ) = 1 \, . \label{eq:pnuijnorm}
\end{equation}

 When $\nu_{i, j} = 0$ ($J^{( r_{i, j} )}_{i, j} = J$)
 for all $(i, j)$ pairs,
  the model becomes the ferromagnetic Potts model.
 When $q = 2$, the model becomes the Edwards-Anderson model
 and is especially called the $\pm J$ Ising model.

 In Ref.~\citen{T},
 it was pointed out that
 a gauge transformation has no effect on thermodynamic quantities.
 To calculating thermodynamic quantities,
 a gauge transformation wherein the transformation is performed by
\begin{equation}
 J^{( r_{i, j} ) }_{i, j} \to J^{( r_{i, j} )}_{i, j} \mu^{q - r_{i, j}}_i
 \mu^{r_{i, j}}_j \, , \quad \lambda_i \to \lambda_i \mu_i  \label{eq:GaugeT} 
\end{equation}
 is used, where $\mu_i = e^{\frac{2 \pi i}{q} \tilde{q}_i}$, and
 $\tilde{q}_i$ is an arbitary valuve for $q_i$.
 This gauge transformation was proposed in Ref.~\citen{NS}.
 By the gauge transformation,
 the Hamiltonian  ${\cal H}$ part becomes ${\cal H} \to {\cal H}$.

By using
 Eqs.~(\ref{eq:Pnuij}), (\ref{eq:pnuijnorm}), and (\ref{eq:Pnuij2}),
 the distribution $P ( \nu_{i, j} )$ is given by \cite{NS}
\begin{equation}
 P ( \nu_{i, j} ) =  A e^{\frac{\beta_{\rm P} }{q}
 \sum_{r_{i, j} = 1}^{q - 1} J^{(r_{i, j})}_{ij} ( \nu_{i, j} )} \, ,
 \label{eq:Pnuij2}
\end{equation}
 where $A$ and $\beta_P$ are respectively 
\begin{eqnarray}
 A &=& \frac{1}{e^{\frac{\beta_P J}{q} (q - 1)}
 + (q - 1) e^{- \frac{\beta_P J}{q}}}
  \, , \label{eq:A} \\
 \beta_P &=& \frac{1}{J}
 \ln \biggl[ p \biggl( \frac{q - 1}{1-p} \biggr)
 \biggr] \, . \label{eq:betaPnuij}
\end{eqnarray}
By performing the gauge transformation,  
 the distribution $P (\nu_{i, j})$ part becomes
\begin{eqnarray}
  \prod_{\langle i, j \rangle} P (\nu_{i, j}) 
  &=& 
 A e^{\frac{\beta_{\rm P} }{q} \sum_{\langle i, j \rangle}
 \sum_{r_{i, j} = 1}^{q - 1} J^{(r_{i, j})}_{i, j} ( \nu_{i, j} )} 
 \nonumber \\  &\to& 
 \frac{A}{q^N} \sum_{\{ \mu_i \}} e^{\frac{\beta_{\rm P} }{q} \sum_{\langle i, j \rangle}
 \sum_{r_{i, j} = 1}^{q - 1}  
 J^{( r_{i, j} )}_{i, j} ( \nu_{i, j} ) \mu^{q - r_{i, j}}_i
 \mu^{r_{i, j}}_j } \, ,  \label{eq:Pnuij3}
\end{eqnarray}
 where $\langle x, y \rangle$ denotes the nearest neighbor pairs conneced by links.

%%%%%%%%%%%%%%%%%%
\section{The Fortuin-Kasteleyn cluster} \label{sec:3}
%%%%%%%%%%%%%%%%%%

The bond for the FK cluster is put between spins
 with probability $P_{\rm FK} (q_i , q_j, \nu_{i, j})$.
 The value of $P_{\rm FK}$ depends on
 the interaction between spins and the states of spins.
 We call the bond the FK bond in this article.
 $P_{\rm FK} (q_i, q_j, \nu_{i, j})$ is given by 
\begin{equation}
 P_{\rm FK} (q_i, q_j, \nu_{i, j}) =
 1 - e^{- \frac{\beta}{q} [
 \sum_{r_{i, j} = 1}^{q - 1} J^{(r_{i, j} ) }_{i, j} (\nu_{i, j} )
 \lambda^{r_{i, j}}_i ( q_i )
 \lambda^{q - r_{i, j}}_j ( q_j ) + J ]} \, , 
 \label{eq:PbondliljJij}
\end{equation}
 where $\beta$ is the inverse temperature, and $\beta = 1 / k_B T$.
 $k_B$ is the Boltzmann constant, and $T$ is the temperature.
 By connecting the FK bonds, the FK clusters are generated.
 In appendix, we will derive Eq.~(\ref{eq:PbondliljJij}).
 By the gauge transformation,
 the $P_{\rm FK}$ part becomes $P_{\rm FK} \to P_{\rm FK}$.

\begin{figure}[t]
\begin{center}
\includegraphics[width=0.65\linewidth]{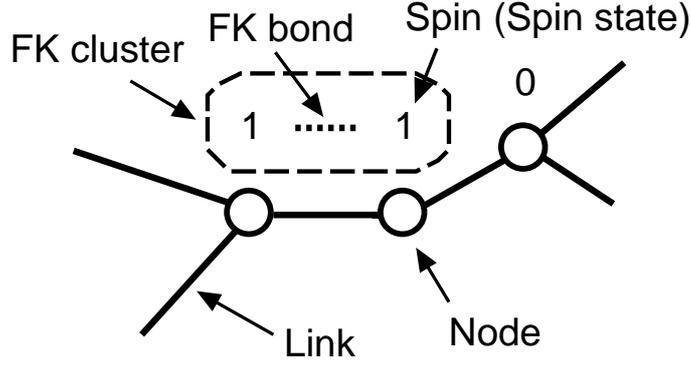}
\end{center}
\caption{
 Network and FK cluster.
 Three nodes, six links, three spins, an FK bond, and an FK cluster are depicted.
 Spins are aligned on each node and are represented by spin states.
 In this picture,
 the states of two spins are one and the state of a spin is zero.
 \label{fig:network-cluster-Potts}
}
\end{figure}
 Figure \ref{fig:network-cluster-Potts} shows a conceptual diagram
 of a network and an FK cluster.
 Three nodes, six links, three spins, an FK bond, and an FK cluster are depicted.
 Spins are aligned on each node and are represented by spin states.

The thermodynamic quantity of the FK bond put between
 the spins on nodes $i$ and $j$,
 $[\langle b_{\rm FK} (i, j) \rangle_T ]_R$, is given by
\begin{equation}
 [\langle b_{\rm FK} (i, j) \rangle_T ]_R =
 [\langle  P_{\rm FK} (q_i, q_j, \nu_{i, j}) \rangle_T ]_R \, , \label{eq:bbondijTR}
\end{equation}
 where $\langle  \, \rangle_T$ is the thermal average,
 and $[ \, ]_R$ is the random configuration average.
  The thermodynamic quantity of the node degree for FK bonds at node $i$,
 $[\langle k_{\rm FK} (i) \rangle_T ]_R$, is given by
\begin{equation}
 [\langle k_{\rm FK} (i) \rangle_T ]_R  
 = [\langle  \sum_{\{ j | b(i, j) = 1\}}
 P_{\rm FK} (q_i, q_j, \nu_{i, j}) \rangle_T ]_R \, . \label{eq:kbondiTR}
\end{equation}
 The thermodynamic quantity of the square of the node degree for FK bonds at node $i$,
 $[\langle k^2_{\rm FK} (i) \rangle_T ]_R$, is given by
\begin{eqnarray}
 & & [\langle k^2_{\rm FK} (i) \rangle_T ]_R \nonumber \\
 &=& [\langle  \sum_{\{ j | b(i, j) = 1\}}
 \sum_{\{ l | b(i, l) = 1\}} P_{\rm FK} (q_i, q_j, \nu_{i, j})
 P_{\rm FK} (q_i, q_l, \nu_{i, l}) (1 - \delta_{j, l} )   
  \nonumber \\
 &+& \sum_{\{ j | b(i, j) = 1\}} P_{\rm FK} (q_i, q_j, \nu_{i, j}) \rangle_T ]_R \, . 
 \label{eq:k2bondiTR}
\end{eqnarray}
 The thermodynamic quantity of the node degree for FK bonds,
 $[\langle k_{\rm FK} \rangle_T ]_R$, is given by
\begin{equation}
 [\langle  k_{\rm FK} \rangle_T ]_R = \frac{1}{N} \sum_i^N
 [\langle  k_{\rm FK} (i) \rangle_T ]_R \, . \label{eq:kbondTR} 
\end{equation}
 The thermodynamic quantity of the square of the node degree for FK bonds,
 $[\langle k^2_{\rm FK} \rangle_T ]_R$, is given by
\begin{equation}
 [\langle  k^2_{\rm FK} \rangle_T ]_R = \frac{1}{N} \sum_i^N
 [\langle  k^2_{\rm FK} (i) \rangle_T ]_R \, . \label{eq:k2bondTR}
\end{equation}

%%%%%%%%%%%%%%%%%%
\section{A criterion for percolation of cluster} \label{sec:4}
%%%%%%%%%%%%%%%%%%

We use a conjectured criterion for deriving the percolation thresholds.
 This criterion is a criterion of the percolation of cluster for spin models
 on the random graphs with arbitary degree distributions,
 and is given by \cite{Y}
\begin{equation}
 [\langle k^2_{\rm bond} \rangle_T ]_R
 \ge 2 [\langle k_{\rm bond} \rangle_T ]_R \, ,
 \label{eq:k2k1bond}
\end{equation}
 where $k_{\rm bond}$ is a quantity for a bond put between spins.
 $k_{\rm bond}$ for the FK bond is $k_{\rm FK}$ for example. 
 Equation~(\ref{eq:k2k1bond}) is given by the inequality when
 the cluster is percolated.
 Equation~(\ref{eq:k2k1bond}) is given by the equality when
 the cluster is at the percolation transition point.

In addition, Eq.~(\ref{eq:k2k1bond}) is true for sufficiently
 large number of nodes in the case that
 the magnitude of the bond does not depend on the degree $k (i)$.

We consider the condition that
 the magnitude of the bond does not depend on the degree $k (i)$.
 We define a variable for the inverse temperature $\beta$ 
 as $\rho (\beta )$.
 We set
\begin{equation}
 0 < \rho (\beta ) \le 1 \, . \label{eq:rhorange}
\end{equation}
 We consider a case that  
 $[\langle b_{\rm bond} (i, j) \rangle_T]_R$,
 $[\langle  k_{\rm bond } (i) \rangle _T]_R$, and 
 $[\langle  k^2_{\rm bond} (i) \rangle_T ]_R$ are respectively written as
\begin{equation}
 [\langle b_{\rm bond} (i, j) \rangle_T ]_R =
 \rho (\beta )  \, , \label{eq:bondij}
\end{equation}
\begin{equation}
 [\langle k_{\rm bond} (i) \rangle_T ]_R =
 \rho (\beta ) \, k (i) \, , \label{eq:k1bondi}
\end{equation}
\begin{eqnarray}
 [\langle k^2_{\rm bond} (i) \rangle_T ]_R &=& 
 \rho^2 ( \beta ) \, k (i) 
 [ k (i) - 1 ]  \nonumber \\
 &+& \rho ( \beta ) \, k (i) \, .
 \label{eq:k2bondi}
\end{eqnarray}
 In the case, it is implied that
 the bias for $k (i)$ does not appear
 in the statistical results of the bonds.
 Therefore, we describe the case that
 $[\langle b_{\rm bond} (i, j) \rangle_T]_R$,
 $[\langle  k_{\rm bond } (i) \rangle _T]_R$, and 
 $[\langle  k^2_{\rm bond} (i) \rangle_T ]_R$ are respectively written as 
 Eqs.~(\ref{eq:bondij}), (\ref{eq:k1bondi}), and (\ref{eq:k2bondi}) as the case 
 that the magnitude of the bond does not depend on $k (i)$.

This criterion is a conjectured criterion and is not exactly derived yet.
 On the other hand, it was confirmed~\cite{Y} that
 this criterion is exact for several extremal points
 when applied to the Edward-Anderson model.
 In this article,
 we do not examine this criterion and
 just apply this criterion to the present system.

%%%%%%%%%%%%%%%%%%
\section{Results} \label{sec:5}
%%%%%%%%%%%%%%%%%%

We will obtain the percolation thresholds of the FK cluster in this section.

By using Eqs.~(\ref{eq:GaugeT}), (\ref{eq:Pnuij3}),
 (\ref{eq:PbondliljJij}), and (\ref{eq:bbondijTR}),
 when $\beta = \beta_P$, 
 the thermodynamic quantity of the FK bond put between
 the spins on nodes $i$ and $j$,
 $[\langle b_{\rm FK} (i, j) \rangle_T ]_R$, is obtained as
\begin{eqnarray}
 & & [\langle b_{\rm FK} (i, j ) \rangle_T ]_R \nonumber \\
 &=& \sum_{ \{ \nu_{l, m} \}}
 \prod_{\langle l, m \rangle} P ( \nu_{l, m})  
  \frac{\sum_{\{ q_l \} } 
 P_{\rm FK} ( q_i, q_j, \nu_{i, j})
 \, e^{ - \beta_P {\cal H} (\{ q_l \}, \{ \nu_{l, m} \})}}
 {\sum_{\{ q_l \} } 
 e^{- \beta_P {\cal H} (\{ q_l \}, \{ \nu_{l, m} \})}} \nonumber \\
  &=& \frac{A^{N_B}}{q^N}  \sum_{ \{ \nu_{l, m} \}}
 \sum_{\{ S_l \} } P_{\rm FK} (q_i, q_j, \nu_{i, j})
 \, e^{- \beta_P {\cal H} (\{ q_l \}, \{ \nu_{l, m} \})} \nonumber \\ 
 &=& \frac{e^{\beta_P J} - 1}{e^{\beta_P J} + q - 1 } \, ,
 \label{eq:bbondij}
\end{eqnarray}
 where $N_B$ is the number of all links, and $N_B = N \langle  k \rangle_N / 2$.
 By using Eqs.~(\ref{eq:GaugeT}), (\ref{eq:Pnuij3}),
 (\ref{eq:PbondliljJij}), and (\ref{eq:kbondiTR}),
 when $\beta = \beta_P$, 
 the thermodynamic quantity of the node degree for FK bonds at node $i$,
 $[\langle k_{\rm FK} (i) \rangle_T ]_R$, is obtained as
\begin{equation}
 [\langle k_{\rm FK} (i) \rangle_T ]_R =
 \biggl( \frac{e^{\beta_P J} - 1}{e^{\beta_P J} + q - 1 } \biggr)
 \, k (i) \, . \label{eq:k1bondi2}
\end{equation}
By using
 Eqs.~(\ref{eq:GaugeT}), (\ref{eq:Pnuij3}), (\ref{eq:PbondliljJij}),
 and (\ref{eq:k2bondiTR}),
 when $\beta = \beta_P$, 
 the thermodynamic quantity of the square of the node degree for FK bonds at node $i$,
 $[\langle k^2_{\rm FK} (i) \rangle_T ]_R$, is obtained as
\begin{eqnarray}
 & & [\langle k^2_{\rm FK} (i) \rangle_T ]_R \nonumber \\
 &=& 
 \biggl( \frac{e^{\beta_P J} - 1}{e^{\beta_P J} + q - 1 } \biggr)^2
  \,  k (i) [ k (i) - 1 ] 
 \nonumber \\
 &+& 
  \biggl( \frac{e^{\beta_P J} - 1}{e^{\beta_P J} + q - 1 } \biggr)
  \, k (i) \, .
 \label{eq:k2bondi2}
\end{eqnarray}

 When we set
\begin{equation}
 \rho ( \beta_P ) = 
 \frac{e^{\beta_P J} - 1}{e^{\beta_P J} + q - 1 }
 \, , \label{eq:rho}
\end{equation}
 Eqs~(\ref{eq:bbondij}),
 (\ref{eq:k1bondi2}), and (\ref{eq:k2bondi2}) are
 respectively formulated as Eqs.~(\ref{eq:bondij}),
 (\ref{eq:k1bondi}), and (\ref{eq:k2bondi}).
 Therefore,
 the magnitude of the bond does not depend on the $k (i)$.
 By using Eqs.~(\ref{eq:kbondTR}), (\ref{eq:k2bondTR}), (\ref{eq:k2k1bond}),
 (\ref{eq:k1bondi2}), and (\ref{eq:k2bondi2}), we obtain
\begin{equation}
 \frac{2 (e^{\beta_P J} - 1)}{2 (e^{\beta_P J} - 1) + q}
 \ge \frac{2 \langle k \rangle_N}{\langle k^2 \rangle_N} \, . \label{eq:bondPc}
\end{equation}
 Equation~(\ref{eq:bondPc}) is given by the inequality when 
 the cluster is percolated.
 Equation~(\ref{eq:bondPc}) is given by the equality when 
 the cluster is at the percolation transition point. 

From Eqs.~(\ref{eq:rhorange}) and (\ref{eq:rho}),
 there is the percolation transition point for $0 < \beta_P \le \infty$.
 From Eq.~(\ref{eq:bondPc}),
 there is the percolation transition point for $0 < a \le 1$.
By using Eqs.~(\ref{eq:betaPnuij}) and (\ref{eq:bondPc}),
 the probability $p$ that 
 the interaction is ferromagnetic is obtained as
\begin{equation}
 p = \frac{2 (1 - a) + a q}{(2 - a) q} \label{eq:pbond}
\end{equation}
 at the percolation transition point.
By using Eqs.~(\ref{eq:betaPnuij}) and (\ref{eq:pbond}), 
 the percolation transition temperature $T_P$ is obtained as
\begin{equation}
 T_P = \frac{J}{k_B \ln \biggl[ \frac{2 ( 1 - a ) + a q}{2 (1 - a)}
  \biggr]}
 \, . \label{eq:TPbond}
\end{equation}

\begin{figure}[t]
\begin{center}
\includegraphics[width=0.95\linewidth]{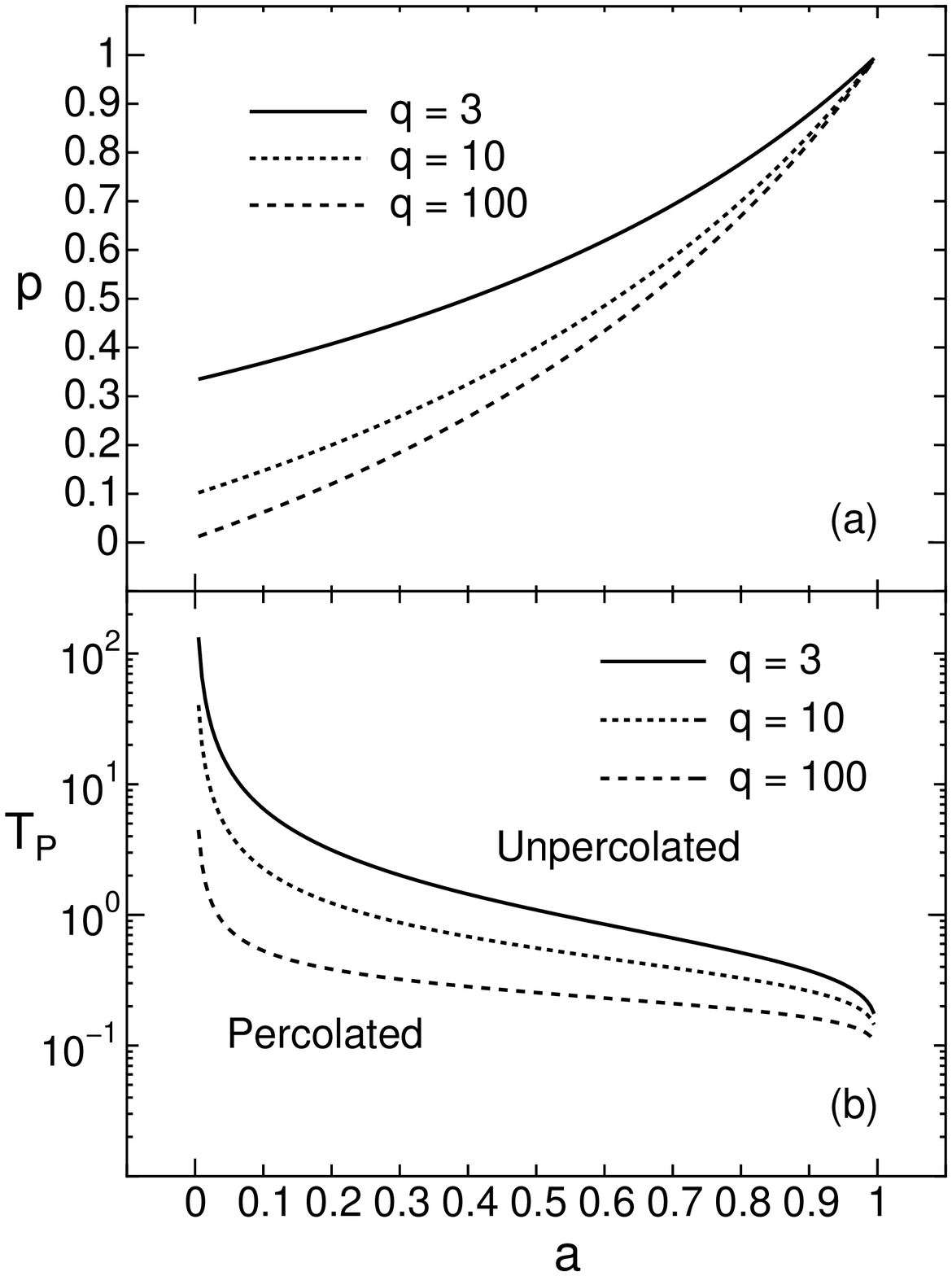}
\end{center}
\caption{
 Percolation thresholds of the FK cluster for the Potts gauge glass model.
 (a) The relation between the aspect $a$ and the probability
 $p$ is shown.
 (b) The relation between the aspect $a$ and the percolation
 transition temperature $T_P$ is shown. 
 The solid line shows the result of $q = 3$,
 the dotted line shows the result of $q = 10$,
 and the short-dashed line shows the result of $q = 100$.
 $J / k_B$ is set to 1.
\label{fig:GPG}
}
\end{figure}
Figure~\ref{fig:GPG} shows
 the percolation thresholds of the FK cluster for the Potts gauge glass model.
 Figure~\ref{fig:GPG}(a) shows
 the relation between the aspect $a$ and the probability $p$.
 Eq.~(\ref{eq:pbond}) is used for showing Fig.~\ref{fig:GPG}(a).
 Figure~\ref{fig:GPG}(b) shows
 the relation between the aspect $a$ and the percolation
 transition temperature $T_P$.
 Eq.~(\ref{eq:TPbond}) is used for showing Fig.~\ref{fig:GPG}(b).
 The solid line shows the result of $q = 3$,
 the dotted line shows the result of $q = 10$,
 and the short-dashed line shows the result of $q = 100$.
 $J / k_B$ is set to $1$.

For the ferromagnetic Potts model on the same network,
 the phase transition temperature $T^{({\rm Ferro})}_C$ is \cite{DGM2}
\begin{equation}
 T^{({\rm Ferro})}_C
  = \frac{J}{k_B \ln \biggl[ \frac{2 ( 1 - a ) + a q}{2 (1 - a)}
    \biggr]} \, .
\end{equation}
 $T_P$ (Eq.~(\ref{eq:TPbond}))
 coincides with 
 $T^{({\rm Ferro})}_C$.

The complete graph is considerd as $a \sim 0$.
 We set
 $\langle k \rangle_N = N - 1$, $\langle k^2 \rangle_N = (N - 1)^2$,
 $a = 2 / (N - 1)$, and $J \to J / \sqrt{N}$.
 From the settings, the model on the network becomes the infinite-range model.
 By using Eq.~(\ref{eq:pbond}), 
 the probability $p^{({\rm IR})}$ that 
 the interaction is ferromagnetic is obtained as
\begin{equation}
 p^{({\rm IR})} = \frac{N - 3 + q}{(N - 2) q} \to \frac{1}{q}  \label{eq:pbond2}
\end{equation}
 for a sufficiently large number of nodes at the percolation transition point.
By using Eq.~(\ref{eq:TPbond}),
 the percolation transition temperature $T^{({\rm IR})}_P$ is obtained as
\begin{equation}
 T^{({\rm IR})}_P =
 \frac{J}{k_B \sqrt{N} \ln ( 1 + \frac{q}{N - 3})} 
 \to \frac{J \sqrt{N}}{k_B q}  \label{eq:TPbond2}
\end{equation}
 for a sufficiently large number of nodes.

%%%%%%%%%%%%%%%%%%
\section{Summary} \label{sec:6}
%%%%%%%%%%%%%%%%%%

In this article,
 the Potts gauge glass model on the random graphs with artibary 
 degree distributions was investigated.

The value of 
 $[\langle b_{\rm FK} (i, j) \rangle_T]_R$,
 $[\langle k_{\rm FK} (i) \rangle_T]_R$, $[\langle k^2_{\rm FK} (i) \rangle_T]_R$,
 $[\langle k_{\rm FK} \rangle_T]_R$, and $[\langle k^2_{\rm FK} \rangle_T]_R$
 on the Nishimori line were shown.
 They are quantities for the FK bonds and are exact even on a finite number of nodes.

It is known that 
 the internal energy and the upper bound of the specific heat 
 are exactly calculated
 on the Nishimori line in the Potts gauge glass model
 without the dependence 
 of the network (lattice) \cite{NS}.
 In this article, 
 it was realized that, as a property for the Nishimori line, 
 the magnitude of the FK bond does not depend on the degree $k (i)$.

In this article,
 we showed the percolation thresholds of the FK cluster.
 It was shown that
 the percolation transition temperature $T_P$ (Eq.~(\ref{eq:TPbond}))
 on the Nishimori line 
 for the Potts gauge glass model on the present network coincides with
 the phase transition temperature $T^{({\rm Ferro})}_C$ \cite{DGM2}
 for the ferromagnetic Potts model on the same network.
 The percolation thresholds of the FK cluster
 for the infinite-range model were also shown.

We used a conjectured criterion Eq.~(\ref{eq:k2k1bond})
 to obtain the percolation thresholds.
 For this criterion, it was confirmed~\cite{Y} that
 this criterion is exact for several extremal points
 when applied to the Edward-Anderson model.
 Therefore, our entire set of results may be exact.

%%%%%%%%%%%%%%%%%%
\section*{Acknowledgment}
%%%%%%%%%%%%%%%%%%

 We would like to thank F. Igl\'oi for useful comments.

%%%%%%%%%%%%%%%%%%
\section*{Appendix: the probabilities for connecting spins}
%%%%%%%%%%%%%%%%%%

We will derive Eq.~(\ref{eq:PbondliljJij})
 according to the method of Kawashima and Gubernatis in Ref.~\citen{KG}.
 We define the weight of two spins on nodes connected by a link
 as $w (q_i, q_j, \nu_{i, j})$. 
 $w (q_i, q_j, \nu_{i, j})$ is given by
\begin{equation}
  w (q_i, q_j, \nu_{i, j}) 
 = \exp \biggl\{ \frac{\beta J}{q}
 \sum_{r_{i, j} = 1}^{q - 1}
 \exp \biggl[ \frac{2 \pi i}{q}
 \biggr( \nu_{i, j} + q_i - q_j \biggl) r_{i, j}
 \biggr] \biggr\} \, . 
 \label{eq:a-1}
\end{equation}
We define the weight $w$ for $\nu_{i, j} + q_i - q_j = 0$
 as $w_{\rm para}$. We obtain
\begin{equation}
 w_{\rm para} (q_i, q_j, \nu_{i, j}) 
 = \exp \biggl[ \frac{\beta J (q - 1) }{q} \biggr] \, . \label{eq:a-2}
\end{equation}
We define the weight $w$ for $\nu_{i, j} + q_i - q_j \ne 0$
 as $w_{\rm anti}$. We obtain
\begin{equation}
 w_{\rm anti} (q_i, q_j, \nu_{i, j})
  = \exp \biggl( - \frac{\beta J}{q} \biggr) \, . \label{eq:a-3}
\end{equation}
 We define the weight of graph for connecting two spins as $w (g_{\rm conn})$.
 We define the weight of graph for disconnecting two spins
 as $w (g_{\rm disc})$.
 We are able to write 
\begin{eqnarray}
 w_{\rm para} (q_i, q_j, \nu_{i, j})
  &=& w (g_{\rm conn}) + w (g_{\rm disc}) \, , \label{eq:a-5} \\
 w_{\rm anti} (q_i, q_j, \nu_{i, j})
 &=& w (g_{\rm disc}) \, .  \label{eq:a-6} 
\end{eqnarray}
 By using Eqs.~(\ref{eq:a-2}), (\ref{eq:a-3}),
 (\ref{eq:a-5}), and (\ref{eq:a-6}),
 we obtain
\begin{eqnarray}
 w (g_{\rm conn}) 
 &=& \exp \biggl[ \frac{\beta J (q - 1) }{q} \biggr] 
 -  \exp \biggl( - \frac{\beta J}{q} \biggr) \, , \label{eq:a-7} \\
 w (g_{\rm disc})
 &=&  \exp \biggl( - \frac{\beta J}{q} \biggr) \, .  \label{eq:a-8}
\end{eqnarray}
 We define the probability of
 connecting two spins for $\nu_{i, j} + q_i - q_j = 0$ as
 $P_{\rm para} (g_{\rm conn})$.
 We define the probability of
 connecting two spins for $\nu_{i, j} + q_i - q_j \ne 0$
 as $P_{\rm anti} (g_{\rm conn})$.
 We are able to write 
\begin{eqnarray} 
 P_{\rm para} (g_{\rm conn})
 &=& \frac{w (g_{\rm conn})}{w (g_{\rm conn}) + w (g_{\rm disc})} \,
 , \label{eq:a-9} \\
 P_{\rm anti} (g_{\rm conn})
 &=& 0 \, . \label{eq:a-10}
\end{eqnarray} 
 By using Eqs.~(\ref{eq:a-7}), (\ref{eq:a-8}), (\ref{eq:a-9}), and (\ref{eq:a-10}), 
 we derive Eq.~(\ref{eq:PbondliljJij}).

\end{document}